\documentclass{elsart}
\usepackage{graphicx}

\begin{document}

\begin{frontmatter}
\title{ Non-rigid shell model and novel correlational effects in atomic and molecular systems}
\author{Yu.D. Panov, A.S. Moskvin}
\address{Department of Theoretical Physics, Ural State University,  620083, Ekaterinburg, Russia}

\begin{abstract}
Direct analytical and numerical calculation show that two-electron atomic
configuration can be unstable with respect to a static or dynamic shift of the
electron shells. This enables  to develop a so called non-rigid shell model for
a partial account of the electron correlations within atomic clusters in
solids. In a framework of this model a correlated state of two-electron
molecular configuration is described by a set of  symmetrized shell shifts
$q^{\gamma}$ similarly to the well known shell model developed for a
description of the lattice dynamics. A set of $q^{\gamma}$-shifts are found
after minimization of the energy functional. We present a number of  the novel
unconventional effects including: i) a correlational mechanism of the local
pairing; ii)  a correlational  (pseudo) Jahn-Teller effect provided by a joint
account of the electron shell shifts and conventional nuclear displacements;
iii) an appearance of the chiral correlational states. The model allows an
introduction of the pseudo-spin formalism and effective "spin-Hamiltonian" for
a description of  the short- and long-range  ordering of non-rigid atomic
backgrounds in crystals. Finally, the model can be readily built in the
conventional band schemes.
\end{abstract}
\end{frontmatter}

\section{ Introduction}

Electronic correlations is one of the fundamental problem in a theory of atoms,
molecules and solid state,  particularly  for the systems with high density of
excited states when a small perturbation can result both in drastic
reconstruction of the energy spectrum and in modification of the ground state
up to formation of a strongly correlated state. As a rule, in such a situation
an appropriate description of the ground state within the bare restricted basis
often requires a lot of configurations or considerable extention of the basis,
and so becomes difficult for practical realization and interpretation. Namely
this situation occurs in atoms, where description of some specific correlation
effects in  terms of Hartree-Fock basis requires a large number of Hartree-Fock
configurations. Such a problem implies a search for  alternative variational
approaches to the electronic structure and energy spectrum.

In this work we develop further  the variational method for the many-electron
atomic clusters  with the trial parameters being the coordinates of the center
of the one-particle atomic orbital \cite{Kozman}. The resulting shift of the
atomic orbital allows to interpret the variation of the electronic density
distribution rather clearly, and the symmetry of a system can be readily used
for construction of the trial many-electron wave function. The shifted
electronic shells in conventional MO-LCAO-scheme with restricted set of the
one-particle states allow to take into account an additional multipole
interactions and to construct  novel states with unique properties. As a whole
the model bears a strong resemblance to the well known shell model widely used
in lattice dynamics.

\section{ Two-electron configuration}

\begin{figure}[th]
\begin{center}
\includegraphics[width=0.4\linewidth]{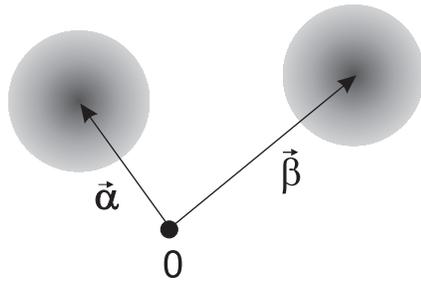}
\caption{ The displacement vectors $\vec{\alpha}$ and $\vec{\beta}$ define the
centers of the one-particle orbitals relatively to the potential center.}
\label{dispvect}
\end{center}
\end{figure}

We consider the problem of two electrons in certain atomic potential to be a
simplest model for manifestation of electronic correlations. The orbital part
of the singlet two-electron wave function formed by the shifted one-particle
orbitals (bi-orbital) can be written as follows:
\begin{equation}
\Psi \left( \vec{r}_1 ,\vec{r}_2 ;\vec{\alpha} ,\vec{\beta} \right) =
\eta^{-1} \left[ \psi (\vec{r}_1 - \vec{\alpha}) \,
\psi(\vec{r}_2 - \vec{\beta}) + \psi(\vec{r}_1 - \vec{\beta}) \,
\psi(\vec{r}_2 - \vec{\alpha}) \right] , \label{WF}
\end{equation}
where $\vec{\alpha}$, $\vec{\beta}$  are the displacement vectors for the
one-particle orbitals (Fig.\ref{dispvect}), $\eta$  the normalization factor.
Below, only the real functions of $s$-type are used as the trial one-particle
states. Then
\begin{equation}
\eta^{2}= 2\,\left( 1+S^{2}\left( \vec{\alpha } ,\vec{\beta} \right) \right) ,
\end{equation}
where $S(\vec{\alpha} ,\vec{\beta})$ is the overlap integral for the
one-particle orbitals. The Hamiltonian of the problem in  atomic units
($\epsilon_0 = \frac{me^4}{\hbar^2} = \frac{e^2}{a_0}$, $a_0 =
\frac{\hbar^2}{me^2}$) is
\begin{equation}
\hat{H} = -\frac{\Delta_1}{2} - \frac{\Delta_2}{2} -
\frac{Z_0}{r_1} - \frac{Z_0}{r_2} +
\frac{1}{\left| \,\vec{r}_1 -\vec{r}_2 \,\right| } .
\end{equation}
The variational procedure is performed with the full energy functional:
\begin{equation}
E \left\{ \Psi \right\} \equiv \left\langle \Psi \,
\left| \,\hat{H}\,\right| \,\Psi \right\rangle =
E\left( \vec{\alpha} ,\vec{\beta} \right) .
\end{equation}
Taking into account the expression (\ref{WF}) we obtain:
\begin{eqnarray}
E( \vec{\alpha} ,\vec{\beta} ) &=&
\frac{1}{1+S^{2}(\vec{\alpha},\vec{\beta})}
[ 2\,t(\vec{\alpha},\vec{\alpha}) -
Z_0 \left( u(\vec{\alpha},\vec{\alpha}) + u(\vec{\beta},\vec{\beta}) \right) +
\label{Energy}\\
&& 2 \, S(\vec{\alpha},\vec{\beta}) \, t(\vec{\alpha},\vec{\beta}) -
2 Z_0 \, S(\vec{\alpha},\vec{\beta}) \, u(\vec{\alpha},\vec{\beta}) +
c(\vec{\alpha},\vec{\beta}) + a(\vec{\alpha},\vec{\beta}) ] ,
\nonumber
\end{eqnarray}
where the following matrix elements are introduced: the one-center integral
$t(\vec{\alpha},\vec{\alpha})$ is the kinetic energy of an electron with
functions of the same center, the two-center integral
$u(\vec{\alpha},\vec{\alpha})$ is the interaction of an electron with the
potential center with functions of the same center, the two-center integral
$t(\vec{\alpha},\vec{\beta})$ is the kinetic energy of an electron with
functions of different centers, the three-center integral
$u(\vec{\alpha},\vec{\beta})$ is the interaction of an electron with the
potential center with functions of different centers,
$c(\vec{\alpha},\vec{\beta})$ and $a(\vec{\alpha},\vec{\beta})$ are the Coulomb
and the exchange parts of the inter-electron interaction. One should note that
$S(\vec{\alpha},\vec{\beta})$, $c(\vec{\alpha},\vec{\beta})$,
$a(\vec{\alpha},\vec{\beta})$ are the two-center integrals.

In further we use the Slater functions with the index $k$ and
effective charge $Z$ as the one-particle atomic orbitals:
\begin{equation}
\psi \left( \vec{r} \right) = N_{Z,k} \, r^k e^{-Zr} , \label{WFSlat}
\end{equation}
where the normalization factor is
\begin{equation}
N_{Z,k} = \sqrt{\frac{Z}{2\pi \, (2k+2)!}} (2Z)^{k+1} .
\end{equation}
The analytic expressions for the matrix elements are presented in Appendix.
Their examination allows to obtain some information about extremal values of
the $\vec{\alpha}$ and $\vec{\beta}$. Introducing
\begin{equation}
\vec{q}_+ =\frac{1}{2} \left( \vec{\alpha} +\vec{\beta} \right),
\vec{q}_- =\frac{1}{2} \left( \vec{\alpha} -\vec{\beta} \right)
\end{equation}
and the coordinate system with center at $\vec{q}_+ = 0$, one can see that only
$u(\vec{\alpha},\vec{\alpha})+u(\vec{\beta},\vec{\beta})$,
$u(\vec{\alpha},\vec{\beta})$ depend on $\vec{q}_+$:
\begin{eqnarray}
u(\vec{\alpha},\vec{\alpha}) + u(\vec{\beta},\vec{\beta}) &=&
\int \frac{d\vec{r}}{\left| \, \vec{r} - \vec{q}_+ \,\right|} \,
\left( \psi^2(\vec{r} - \vec{q}_{-}) + \psi^2(\vec{r} + \vec{q}_{-}) \right), \\
u(\vec{\alpha},\vec{\beta}) &=&
\int \frac{d\vec{r}}{\left| \, \vec{r} - \vec{q}_{+} \,\right|} \,
\psi(\vec{r} - \vec{q}_{-}) \, \psi(\vec{r} + \vec{q}_{-}).
\nonumber
\end{eqnarray}
These quantities are invariants with respect to inversion in the displacement
vector space, hence $\vec{q}_+ = 0$ is a critical point in $\vec{q}_+$-space.
The surface $E\{\Psi\}=const$ in $\vec{q}_+$-space is a sphere for the
one-particle $s$-functions , and the point $\vec{q}_+ = 0$ is a minimum or a
maximum. In general, for an arbitrary angular dependence of the one-particle
functions $\psi$ the point $\vec{q}_+ = 0$ can be also a saddle point.

It is rather clear, why the point $\vec{q}_+ = 0$ appears to be critical: at
the given value of the interaction with the potential center, the configuration
with $\vec{\alpha} = -\vec{\beta}$ can minimize inter-electron repulsion. This
is confirmed by results of numerical minimization of the full energy functional
for the $1s^2$-configuration ($k=0$) which are listed in the Table 1 and 2.
Thus, the following function can be defined
\begin{equation}
\Psi(\vec{r}_1,\vec{r}_2 ; \vec{q}) =
\eta^{-1} \left[ \psi(\vec{r}_1 - \vec{q}) \,
\psi(\vec{r}_2 + \vec{q}) + \psi(\vec{r}_1 + \vec{q}) \,
\psi(\vec{r}_2 - \vec{q}) \right] , \label{WFQ}
\end{equation}
that leads to the reduction of the number of trial parameters. The full energy
functional with functions $\psi(r)$ (\ref{WFSlat}) depends only on the
$q=|\vec{q}|$ , but not on the direction of $\vec{q}$.

The value $Z$ in the calculations mentioned above was a free parameter. This
parameter provides an additional mechanism of the electronic density
redistribution along with the electronic shell shift. For the isolated atom it
is naturally to assume that $Z$ is also the variational parameter, because in
this situation the atomic potential is the only mechanism of the electronic
density redistribution:
\begin{equation}
E \left\{ \Psi \right\} = E(q,Z) \equiv
E(\vec{\alpha} = \vec{q},\vec{\beta} = -\vec{q} ; Z)
\end{equation}

It should be noted that in a crystal the functions with certain "atomic" value
$Z$ which minimizes the energy of the isolated atom can form the strongly
correlated state like shifted electronic shell state for the minimization of
given crystal potential characterized by the parameter $Z_0$.

\section{ Expansion of the full energy functional}

The function (\ref{WFQ}) possess the following property: it has no linear term
in the $q$-expansion at $q=0$. As
\begin{equation}
\frac{ \partial \psi( \vec{r}-\vec{q} ) }{\partial q} |_{q=0} =
- \frac{ \partial \psi( \vec{r}+\vec{q} ) }{\partial q} |_{q=0} ,
\end{equation}
the first derivative of the function (\ref{WFQ}) at $q=0$ turns to zero. Hence,
the $E(q,Z)$ with the functions (\ref{WFQ}) has an extremum at $q=0$,  which
type is defined by sign of the quadratic term $E^{(2)}$ in the $q$-expansion of
the $E(q)$:
\begin{equation}
E(q) \approx E^{(0)} + E^{(2)} q^2 .
\end{equation}
For the functions (\ref{WFSlat}) the full energy functional doesn't depend on
the direction of $\vec{q}$, and therefore we consider  shifts to be  directed
along $z$-axis. The $q$-expansion of the function (\ref{WFQ}) can be written in
the following form:
\begin{equation}
\psi(\vec{r} - \vec{q}) \approx
a(\vec{r}) - b(\vec{r}) \, q + \frac12 c(\vec{r}) \, q^2 ,
\end{equation}
where
\begin{eqnarray}
a(\vec{r}) &=& \psi(\vec{r} - \vec{q}) |_{q=0} =
N_{Z,k} \, r^k e^{-Zr} , \\
b(\vec{r}) &=&
\frac{\partial \psi(\vec{r} - \vec{q})}{\partial z} |_{q=0} =
N_{Z,k} \left( k \, r^{k-1} - Z \, r^k \right) \, e^{-Zr} \cos \theta , \\
c(\vec{r}) &=&
\frac{\partial^2 \psi(\vec{r} - \vec{q})}{\partial^2 z} | _{q=0} =
N_{Z,k} [ k \, r^{k-2} - Z \, r^{k-1} + \\
&& + \left( k (k-2) \, r^{k-2} - Z(2k-1) \, r^{k-1} + Z^2 r^k \right)
\cos^2 \theta ] \, e^{-Zr} ,
\nonumber
\end{eqnarray}
with the angle $\theta$ counted out from $z$-axis. The $q$-expansion of the
matrix elements up to quadratic terms  and the expressions for $E^{(0)}$ and
$E^{(2)}$ are listed in the Table 3.

From the expression for $E^{(2)}$ the criterion of the non-zero shift of the
electronic shell can be obtained for the $1s$-function ($k=0$) as the
one-particle state. The shift is not  zero, if $Z > Z_0 - \frac{3}{16} $; in
opposite case there is no displacement of the electronic shell from the
potential center. This result is entirely compatible with our numerical
calculations presented in Table 2. The minimum of $E(Z,q)$ for $k=0$ is
obtained for $q_{min}=0$ and $Z_{min} = Z_0 - \frac{5}{16}$, that  agrees with
the well-known result in the helium atom theory \cite{Landau}.

From the other hand, the full energy functional with functions (\ref{WFQ})
formed from the one-particle $ns$-states (\ref{WFSlat}) with $k \neq 0$ has the
minimum at $q \neq 0$ for any values $Z_0$ and $Z$ . It means that for these
states  the non-zero shift of the electronic shell always takes place.

The expression for $E^{(0)}$ allows to find the value of the effective charge
$Z$ that minimizes full energy of the $ns^2$-configuration at $q=0$:
\begin{equation}
Z_{min} = Z_0 \frac{2k+1}{k+1} -
\left( \frac{2k+1}{2k+2} -
\frac{(2k+1)(4k+3)!}{2^{4k+3} \left[(2k+2)!\right]^2} \right) .
\end{equation}
At $k \rightarrow \infty$ this expression tends to $Z^{\infty}_{min} = 2 Z_0 - 1$.

Origin of the different behaviour of the electronic shells with $k=0$ and $k
\neq 0$ can be readily understood from the data listed in Table 3. The
principal difference of the one-particle function with $k \neq 0$ from that of
with $k=0$ is that the function with $k \neq 0$ turns into zero at $r=0$
(Fig.2). The energy of interaction of electrons with the potential center
appears to be  most sensitive to the electron density on nucleus. Its
displacement leads to  strong increase of the full energy of the system in the
case of wave function with $k=0$, and, from the other hand, zero density at
$r=0$ for the function with $k \neq 0$ provides specific "softness" of this
part of interaction that is revealed in zero value of quadratic term in
expansion of $u(\vec{q},-\vec{q})$. The expressions for matrix element
$t(\vec{q},-\vec{q})$ describing relative motion of the electron shells are
different at $k=0$ and $k \neq 0$ whereas the overlap integrals
($S(\vec{q},-\vec{q})$, $c(\vec{q},-\vec{q})$, $a(\vec{q},-\vec{q})$) have the
same structure for the both cases. Finally, it can be concluded that the gain
in the electron-electron interaction at $k \neq 0$ with displacement of the
electronic shells hasn't been compensated by the loss in the energy of
interaction with the potential center as in the case $k=0$.

An appearance of the electronic density on the potential center also  implies
the gain in energy. This makes possible to determine the most favorable
directions of displacement of the electronic shells in the case of the
anisotropic wave function with node at $r=0$. So, for $p_z$-orbital it has to
be $z$-axis direction, for $d_{x^2-y^2}$-orbital the critical directions are
along $x$- and $y$-axis, and so on. Finally,  for the $k \neq 0$ orbitals one
should expect a well developed non-trivial $q \neq 0$ minimum at the energy
surface $E(q,Z)$. This conclusion appears to be compatible with the results of
numerical minimization of the full energy functional for $np^2$-configuration
\cite{Moskvin1997}.

\section{ The form of the full energy functional $E(q,Z)$}

The general expressions for the matrix elements in case of the electronic
configuration with $\vec{\alpha}=\vec{q}$ and $\vec{\beta}=-\vec{q}$ can be
obtained from those listed in Table 1 at $\rho=2Zq$, $\tilde{\alpha}=Zq$ and
with transition to the limit $\xi \rightarrow 0$, $\eta \rightarrow 0$ in
expression for $u(\vec{\alpha},\vec{\beta})$. In terms of $\rho=2Zq$ the latter
can be written as:
\begin{equation}
u(\vec{q},-\vec{q}) =
Z \frac{\rho^{2k+2}}{(2k+2)!}
\sum_{s=0}^{k+1}(4s+1) \, b_0^{(s)} \, \Sigma_{s,0}^{(k+1)}(\rho) .
\end{equation}
At $k=0$ the matrix elements coincide with the well-known fundamental results
from the atomic theory \cite{Sommerfeld,Sugiura1927}.

The results of minimization of $E(q,Z)$ for a number of the lowest values of
$k$ ($k=0,...4$) are listed in  Table 4; the full energy functional as a
function of $q$ at $Z=Z^k_{min}$ is shown in Fig.3. As it was mentioned above,
for $k \neq 0$ the displacement of the electronic shells takes place. The gain
in energy is $0.22 \div 0.36 \, a.u.$, and the value of displacement is $0.34
\div 0.76 \, a.u.$ for $k=1,...4$. The $k$-dependence  of the global minimum
can be explained by the character of the electronic density distribution at
different $k$. The Fig.2 shows that  the increasing  in $k$ is accompanied by
the  lower values of the electronic density near $r=0$. The gain in energy of a
system can be provided by decreasing of electron-electron interaction
(decreasing of positive contribution from $t(\vec{q},-\vec{q})$,
$c(\vec{q},-\vec{q})$, $a(\vec{q},-\vec{q})$) and by increasing of the
interaction with the potential center (increasing of negative contribution from
$u(\vec{q},\vec{q})$, $u(\vec{q},-\vec{q})$). The rising of $q_{min}$ with the
index $k$ is associated with the delocalization $\psi_k (r)$; the lower value
of the overlap and, therefore, the lower interaction of the electrons are
obtained with rising of $q$. Another mechanism providing the overlap decrease
is in increasing of $Z$, that leads also to a localization of $\psi(r)$ (see
Fig.4). With increasing $Z$ the negative contribution from interaction with the
potential center and the positive ones from kinetic energy are increased
simultaneously resulting in certain compromise value of $Z$. This value
increases with $k$ due to a delocalization of $\psi(r)$ with $k$. In Fig.5  the
value $q_{min}$ as function of  $Z$ is also shown.

\section{ The dynamic shifts  of the electronic shells}

The distribution of the electronic density at $k=1$ with and without shifts is
shown in Fig.6; in the both cases the wave functions providing the minimum of
the full energy functional are used. The symmetry of the electronic density
distribution with the shifts ($C_{\infty h}$) breaks the initial spherical
symmetry, which can be restored with taking into account  the energy equivalent
configurations with various directions of the displacement vectors.

The full energy functional can have the continuum of the equivalent minima in
the displacement vector space. In this sense the system has the variational
degeneracy. The existence itself, the form and other parameters of the minima
continuum depend on the one-particle states and the parameters of the
potential. In the case of the $ns^2$-configuration mentioned above, only the
value of difference of the displacement vectors of the one-particle states is
fixed $|\vec{\rho}|=|\vec{\alpha}-\vec{\beta}|=2q$ (with
$\vec{\alpha}=\vec{q}$, $\vec{\beta}=-\vec{q}$), so the continuum is a sphere
in the $\vec{q}$ space that restores the initial spherical symmetry of the
system. By analogy with the description of collective motion in nuclei
\cite{Griffin1967}, one can form the linear combinations with a help of the
bi-orbitals $\Psi$ which have  different vectors $\vec{q}_{min}$:
\begin{equation}
\tilde{\Psi}_f \left( \vec{r}_1,\vec{r}_2 \right) =
\int \Psi \left( \vec{r}_1,\vec{r}_2 ; \vec{q} ,-\vec{q} \right)
f \left( \Omega \right) d\Omega , \label{WFSpher}
\end{equation}
where integration is performed on a sphere in the $\vec{q}$ space. Such a
linear combination can provide the lower energy due to the "off-diagonal" in
$\vec{q}$ matrix elements of the full energy functional which take account of
the "interaction" of bi-orbitals. The variational procedure with the functions
(\ref{WFSpher}) yields an integral equation for the function $f \left( \Omega
\right)$:
\begin{equation}
\int d\Omega f\left( \Omega \right)
\left[ K\left( \vec{q} ,\vec{q}^{\prime} \right) -
E\cdot I\left( \vec{q} ,\vec{q}^{\prime} \right) \right]  = 0 , \label{IntEqv}
\end{equation}
where
$$
    K (\vec{q},\vec{q}^{\prime}) = \langle \Psi(\vec{q},-\vec{q}) \, |
    \,\hat{H} \,| \, \Psi(\vec{q}^{\prime},-\vec{q}^{\prime}) \rangle
$$
$$
    I(\vec{q},\vec{q}^{\prime}) = \langle \Psi(\vec{q},-\vec{q}) |
    \Psi(\vec{q}^{\prime},-\vec{q}^{\prime}) \rangle .
$$
With account of the symmetry of the $ns^2$-configuration \cite{Griffin1967} the
following solutions of (\ref{IntEqv}) can be written:
\begin{equation}
f \left( \Omega \right) = Y_{LM}(\theta,\varphi) .
\end{equation}
In other words, for the $ns^2$-configuration with shifted shells the set of orthogonal
states can be introduced:
\begin{equation}
\tilde{\Psi}_{LM} =
N_{LM} \int Y_{LM} \left( \Omega \right) \Psi \left( \vec{q} ,-\vec{q} \right) d\Omega ,
\end{equation}
the terms of which  transform according to irreducible representations of the
rotation group. These states could be called the dynamic ones, as they can
result in the correlational contribution to orbital current. The spectrum of
these states can be not similar to that of the space rotator. Note that for
such states with the dynamic shifts of the electronic shells one might expect
the anomalously large values of the electric (dipole, quadrupole) or magnetic
susceptibilities, and these values reflect the electronic correlation effects.

\section{ The MO-LCAO-scheme with the shifted atomic orbitals}

Introduction of the shifted one-electron atomic orbitals implies   the
generalization of the conventional MO-LCAO-scheme \cite{Preprint}. Instead of
standart set of the molecular orbitals (MO) $\varphi_{\Gamma_0
\gamma_0}(\vec{r},0)$ being the symmetrized combination of the atomic functions
centered in the points of the equilibrium nuclei positions ($\vec{q}_{\Gamma
\gamma}=0$), the new set of the shifted MO:
\begin{equation}
\varphi_{\Gamma_0 \gamma_0} \left( \vec{r},\vec{q}_{\Gamma \gamma} \right) =
\hat{T}_{q_{\Gamma \gamma }}
\varphi_{\Gamma_0 \gamma_0} \left(\vec{r},0\right) , \label{MO}
\end{equation}
should be introduced, where $\vec{q}_{\Gamma \gamma}$ is the symmetrized
coordinate of the atomic orbital displacements in cluster, $\hat{T}_{q_{\Gamma
\gamma }}$  the operator of the symmetrized  displacement. Such an approach is
the natural generalization of the shifted electronic shells model for the
many-atomic cluster. The symmetry group of the wave function (\ref{MO}) is an
intersection of the group germs of the $\Gamma_0$ and $G$. In contrast with the
symmetrized coordinates of the cluster vibrations, the vector $\vec{q}_{\Gamma
\gamma}$ is fixed and defines the certain distorted distribution of the
electronic density. If $\Gamma \neq A_1$, then the function (\ref{MO}) doesn't
possess the proper transformation properties or, in other words, doesn't belong
to certain irreducible representation of the symmetry group for the undistorted
cluster. This situation appears to be  quite similar to the case of a single
center, where the single electron displacement reduces the symmetry of system
to the minimal one (axial).

Supposing that other things being equal the configuration minimizing  the
inter-electron interaction has the lowest energy, we can introduce the
following two-particle wave function
\begin{equation}
    \Psi_{\Gamma_0 \gamma_0;\Gamma \gamma} \! \left(
    \vec{r}_1,\vec{r}_2;\vec{q}_{\tilde{\Gamma} \tilde{\gamma}} \right)
    = N \left( 1 \pm \hat{P}_{12} \right)
    \hat{T}_{q_{\tilde{\Gamma} \tilde{\gamma}}}^{(1)}
    \hat{T}_{-q_{\tilde{\Gamma} \tilde{\gamma}}}^{(2)}
    \varphi_{\Gamma_0 \gamma_0} \! \left( \vec{r}_1 ,0 \right)
    \varphi_{\Gamma_0 \gamma_0} \! \left( \vec{r}_2 ,0 \right)  , \label{MO2}
\end{equation}
where $N$ is the normalization factor, $\hat{P}_{12}$ is the electron
permutation operator, $\hat{T}_{q_{\tilde{\Gamma} \tilde{\gamma}}}^{(i)}$ is
the operator of the symmetrized displacement $q_{\tilde{\Gamma}
\tilde{\gamma}}$ which transforms accordingly to irreducible representation
$\tilde{\Gamma} \tilde{\gamma}$ and operats in the space of the $i$th electron
coordinates. The upper sign corresponds to the singlet wave function, the lower
sign to the triplet one. The transformational properties $\Gamma \gamma$ of the
two-particle wave function (\ref{MO2}) are defined by $\Gamma \gamma = \Gamma_0
\gamma_0 \times \lbrack \tilde{\Gamma} \tilde{\gamma}]^2$ for the singlet, and
$\Gamma \gamma = \Gamma_0 \gamma _{0} \times \tilde{\Gamma} \tilde{\gamma}$ for
the triplet.

\section{ Electronic Jahn-Teller effect}

Non-rigid shell model gives a simple and obvious example of a local pairing
within the two-electron $ns^2$-like configurations to be a result of the
correlation effects. The local pairing is promoted by the presence of a
strongly polarized shell, as well as the orbital degeneracy or quasi-degeneracy
within valent states (for simplicity, $d$-states) through the electric
multipole $s$-$d$-interaction described by the effective "vibronic-like"
Hamiltonian
$$
V^{sd}=\sum _{\gamma}B_{\gamma}\langle \hat V^{\gamma}q^{\gamma}\rangle,
$$
where the $\hat V^{\gamma}$ operator works within $d$-manifold, the
$B_{\gamma}$ are "vibronic-like" parameters. This interaction can result in a
purely electronic Jahn-Teller effect.

In general, one has  to take account of the atomic displacements $Q^{\gamma}$
modes and their interaction with electronic $q^{\gamma}$ shifts:
$$
V_{qQ}=\sum _{\gamma}b_{\gamma}\langle Q^{\gamma}q^{\gamma}\rangle.
$$
This results in a complicated multi-mode Jahn-Teller effect with a
correlational hybridization at the $s$-, $d$-electron modes and the local
structural modes.

This system will have all anomalous properties of a Jahn-Teller center, in
particular, large values of the low-frequency polarizability. It appears that
within the non-rigid shell model the completely filled electron shells do not
quenched and can reveal many peculiarities similar to the nonfilled shells. The
magnitudes of the shell $q^{\gamma}$ shifts are correlation parameters which
may be found by minimizing the energy functional $E(q)$. The quantity
$\Delta=E(0)-E(q_0)$ determines the pairing energy, i.e., the local boson
binding energy.

Non-rigid shell model can be considered to be a generalization of the well
known shell model of the lattice dynamics and of the non-rigid anionic
background model by J.E.Hirsch et al. \cite{Hirsch}. In particular, a
correlational pseudospin formalism can be successfully applied for a
description of the valent states for the atomic systems with a correlational
near degeneracy. Finally, we would like to conjecture the possible importance
of the non-rigid shells correlation effects for a local pairing in copper
oxides.

\section{Non-rigid shell model and hyperfine coupling}

A correlational shift of the one-electron shells could result in a considerable
renormalization of the hyperfine coupling both for the nominally $ns$- and
non-$s$-orbitals. Here we consider only two effects:

1) an appearance of the effective non-$s$-contribution to contact hyperfine
coupling, and

2) an appearance of the effective nuclear quadrupole interactions for the
nominally $ns$-electrons.

Firstly, a shift of the one-electron non-$s$-shells for two-electron
configuration implies an emergence of  the effective electron density on the
nucleus and could be detected in nuclear resonance, first of all in anomalous
isotope (chemical) shift. In other words, the bare $p$-, $d$-, ... electrons
within the non-rigid shell configurations appear to be involved in the contact
hyperfine coupling.

Secondly, a shift of the one-electron shells for two-electron configuration
results in a modification of the nuclear quadrupole coupling due to change in
the electric field gradient. Moreover, breaking the spherical symmetry of the
electron density distribution within the $ns^2$ configurations with the shifted
shells leads to an appearance of the electric field gradient on the nucleus:
$$
V_{ij} \propto q_{i}q_{j}-\frac{1}{3}{\vec q}^{2}\delta _{ij}.
$$
The  effects under consideration could provide real opportunities for a
detection of the correlational shift of the one-electron shells with the help
of various nuclear methods.

\section{Conclusion}

In a framework of the non-rigid shell model a correlated state of two-electron
molecular configuration is described by a set of  symmetrized  shell shifts
$q^{\gamma}$ similarly to the well known shell model developed for a
description of the lattice dynamics. Such a state could appear even for the
completely filled shells thus resulting in a non-rigid atomic background with
internal degrees of freedom. Contrary to conventional approach this background
in common has nonzero electric and magnetic multipole moments. Non-rigid
shell/background model results in a number of  the novel unconventional effects
including: i) a correlational mechanism of the local pairing; ii)  a
correlational  (pseudo) Jahn-Teller effect provided by a joint account of the
electron shell shifts and conventional nuclear displacements; iii) an
appearance of the  correlational current states. The model  allows an
introduction of the pseudo-spin formalism and effective "spin-Hamiltonian" for
a description of  the short- and long-range  ordering of the non-rigid atomic
backgrounds in crystals. Finally, the model can be readily built in the
conventional band schemes.

\newpage

\begin{center}
{\Large Appendix}
\end{center}

Here an expressions of the matrix elements with a single-particle state
$$
\psi_k \left( \vec{r} \right)  = N_{Z,k}\,r^{k}\,e^{-Zr} \;,\quad \mbox{ where
}
 N_{Z,k} = \frac{(2Z)^{k+\frac{3}{2}}}{\sqrt{4\pi \,( 2k+2)! }}
$$
are presented.

The matrix element of kinetic energy of  electron with functions of the same
center is:
\begin{eqnarray}
    t \! \left( \vec\alpha,\vec\alpha \right) =
    \int d\vec{r}\: \psi_k \! \left(\vec{r} \right) \left( -\frac{\Delta}{2} \right)
    \psi_k \! \left( \vec{r} \right)
    =\frac{Z^2}{2(2k+1)}
\end{eqnarray}

\medskip

The matrix element of interaction of  electron with the potential center with
functions of the same center is:
\begin{eqnarray}
     u \! \left( \vec\alpha,\vec\alpha \right)
     &=& \int \frac{d
     \vec{r}}{r}\,\psi_k^2 \! \left( \vec{r}-\vec\alpha \right) \\
    &=& Z \, \left[\frac{1}{\tilde\alpha}-\frac{e^{-2\tilde\alpha}}{\tilde\alpha}
     \sum\limits_{l=0}^{2k+1} \left( 1-\frac{l}{2k+2} \right)
     \frac{\left( 2 \tilde\alpha \right)^l}{l!} \right] \;,  \nonumber
\end{eqnarray}
where $\tilde\alpha = Z \alpha$.

\medskip

The overlap integral for the one-particle orbitals is:
\begin{eqnarray}
    S \! \left( \vec\alpha,\vec\beta \right)
    = \int d\vec{r} \, \psi_k \!\left(\vec{r}-\vec\alpha \right) \,
    \psi_k \!\left( \vec{r}-\vec{\beta} \right)
    = \frac{e^{-\rho}}{(2k+2)!} \sum\limits_{s=0}^{2k+2} C_s^{(k,k)}\,\rho^s \;,
\end{eqnarray}
where $\rho = Z |\vec\alpha-\vec\beta|$,
\begin{eqnarray}
    && A_j^{(n,n')} = \sum\limits_{l=\max\{0,2j-n-1\}}^{\min\{2j,n'+1\}} (-1)^l
    {n+1 \choose 2j-1} {n'+1 \choose l} \;,\;\; {a \choose b} =
    \frac{a!}{b!(a-b)!}\;; \nonumber
\\
&&C_s^{(n,n')} = \sum\limits_{j=0}^{\left[\frac{s}{2}\right]}
  \frac{A_j^{(n,n')}}{2j+1} \frac{(n+n'+2-2j)!}{(s-2j)!} \;,\;
  \mbox{ where } \left[\frac{s}{2}\right] \mbox{-- integer of } s/2 \;, \nonumber
\\
&&\mbox{defined by }
  \frac{2^{n+n'+3}}{4\pi}
  \int d\vec{x} |\vec{x}-\vec\alpha|^n
  e^{-|\vec{x}-\vec\alpha|} |\vec{x}-\vec\beta|^{n'} e^{-|\vec{x}-\vec\beta|} =
  e^{-\rho} \sum\limits_{s=0}^{n+n'+2} \rho^s C_s^{(n,n')}
\;. \nonumber
\end{eqnarray}

\medskip

The matrix element of kinetic energy of  electron with functions of different
centers is:
\begin{eqnarray}
    t \!\left( \vec\alpha,\vec\beta \right)
    &=& \int d\vec{r}\,\psi_k \!\left(
    \vec{r}-\vec\alpha \right) \left( -\frac{\Delta}{2} \right) \psi_k \!\left(
    \vec{r}-\vec\beta \right) \\
    &=& -Z^2 \frac{e^{-\rho }}{2(2k+2)!}
   \Biggl[ \sum\limits_{s=0}^{2k+2} \rho^s C_s^{(k,k)}
   - 4(k+1) \sum\limits_{s=0}^{2k+1} \rho^s C_s^{(k,k-1)} \nonumber\\
   && +\; 4k(k+1) \sum\limits_{s=0}^{2k} \rho^s C_s^{(k,k-2)}\Biggl] \;,\nonumber
\end{eqnarray}

\medskip

The matrix element of Coulomb part of  inter-electron interaction is:
\begin{eqnarray}
    c \!\left( \vec\alpha,\vec\beta \right)
    &=& \int \frac{d \vec{r}_1 d\vec{r}_2}{\left| \, \vec{r}_1-\vec{r}_2 \,\right| } \,
    \psi_k^2 \!\left( \vec{r}_1-\vec\alpha \right) \,
    \psi_k^2 \!\left( \vec{r}_2-\vec\beta \right) \\
    &=& Z \, \Biggl[ \frac{1}{\rho} -\frac{e^{-2\rho}}{(2k+2) \rho}
    \sum\limits_{l=0}^{2k+1} \rho^l \frac{2^l (2k+2-l)}{l!} \nonumber\\
    && - \; \frac{e^{-2\rho}}{2^{2k+2} (k+1)(2k+2)!}
    \sum\limits_{l=0}^{4k+2} \rho^l G_l^{(k)} \Biggl] \;, \nonumber
\end{eqnarray}
where
$$
  G_l^{(k)} = 2^l \sum\limits_{j=\max \left\{0,l-2k-1\right\}}^{2k+1}
  \frac{(2k+2-j)}{2^j j!} \; C_l^{(2k,j-1)} \;.
$$

\medskip

The matrix element of exchange part of  inter-electron interaction is:
\begin{eqnarray}
&& a \!\left( \vec\alpha,\vec\beta \right) =
  \int \frac{d \vec{r}_1 d\vec{r}_2}{\left| \, \vec{r}_1-\vec{r}_2 \,\right| } \,
  \psi_k \!\left( \vec{r}_1-\vec\alpha \right) \,
  \psi_k \!\left( \vec{r}_2-\vec\beta \right)
  \psi_k \!\left( \vec{r}_1-\vec\beta \right) \,
  \psi_k \!\left( \vec{r}_2-\vec\alpha \right) \\
  && = Z \frac{\rho^{4k+5}}{[(2k+2)!]^2}
  \sum\limits_{s=0}^{k+1} (4s+1)
  \sum\limits_{n=0}^{2k+2} \rho^n F_{s,n}^{(k+1)}(\rho)
  \left\{ e^{-\rho } \Sigma_{s,0}^{(k+1)}(\rho)
  -\Sigma_{s,n}^{(k+1)}(2\rho) \right\} \;,\nonumber
\end{eqnarray}
where
\begin{eqnarray}
  && F_{s,n}^{(m)}(\rho) =
    \frac{1}{n!} \sum\limits_{j=\frac{n+1}{2}}^m B_j^{(m,s)}
    \frac{(2j)!}{\rho^{2j+1}} \; ; \;\;
    B_j^{(m,s)} = \sum\limits_{r=\max\{0,j-s\}}^{\min\{j,m-s\}}
    a_r^{(m,s)} b_{j-r}^{(s)} \; ; \nonumber\\
  && \Sigma_{s,n}^{(m)}(x) = \tilde\Sigma_{s,n}^{(m)}(x)
    -\frac{e^{-x}}{x^{n+2}} \tilde{\tilde\Sigma}_{s,n}^{(m)}(x,1) \; ; \;\; \nonumber\\
  && \tilde{\tilde\Sigma}_{s,n}^{(m)}(x,\xi)
    =\sum\limits_{i=0}^{m-1} \frac{(2i+n+1)!}{x^{2i}}
    \tilde{D}_i^{(m,s)} \sum\limits_{t=0}^{2i+n+1} \frac{(x\xi)^t}{t!} \; ; \nonumber\\
  && \tilde\Sigma_{s,n}^{(m)}(x) = \frac{e^{-x}}{2x} \sum\limits_{i=0}^{m-s} a_i^{(m,s)}
     \times \nonumber\\
  && \times \sum\limits_{l=0}^{2s+2i+n} \frac{\sigma_{2i+n,l}^{(s)}(1)}{x^l}
    \left[ \ln 2 \gamma x - S_{0,l} - ({-}1)^{l+n} e^{2x} Ei(-2x) + ({-}1)^{l+n}
    \sum\limits_{h=0}^{l-1} \frac{( 2x)^h}{h!} S_{h,l} \right] \;, \nonumber\\
  && \tilde{D}_i^{(m,s)} =
    \sum\limits_{l=\max\{0,i-s+1\}}^{\min\{i,m-s\}} a_l^{(m,s)} D_{i-l}^{(s)}\; ; \;\;\;
    \sigma_{t,l}^{(s)}(\xi) = \sum\limits_{r=\max\{0,\frac{l-t+1}{2}\}}^s
    b_r^{(s)} \frac{(2r+t)!}{(2r+t-l)!} \xi^{2r+t} \; ;  \nonumber\\
  && S_{h,l} = \sum\limits_{t=h+1}^l \frac1t \; ; \;\;
    a_l^{(m,s)} = (-1)^{m-l} {m \choose l}
    \frac{2^{2s+2}(2m-2l)!(m-l+s+1)!}{(m-l-s)!(2m-2l+2s+2)!}\;\;
    \nonumber\\
  && \mbox{-- the coefficients in }  \sum\limits_{l=0}^m a_l^{(m,s)} x^l =
    \int\limits_{-1}^1 \left( x^{2}-t^{2} \right)^m P_{2s}(t)dt \; ; \nonumber\\
  && \mbox{ the coefficients } b_l^{(s)} \mbox{ and } D_{l}^{\left( s\right) }
    \mbox{ define the Legendre polynomial $P_{2s}(x)$:} \nonumber\\
  && P_{2s}(x) = \sum\limits_{l=0}^{s} b_l^{(s)} x^{2l}\; , \;\; b_l^{(s)} =
    \frac{(-1)^{s-l}}{2^{2s}} \frac{(2s+2l)!}{(s-l)!(s+l)!(2l)!} \;, \nonumber\\
  && \mbox{ and the Legendre polynomial of second type $Q_{2s}(x)$:}\nonumber\\
  && Q_{2s}(x) = \frac12 \ln\frac{x+1}{x-1} P_{2s}(x)
    -\sum\limits_{l=0}^{s-1} D_l^{(s)} x^{2l+1} \; ; \nonumber\\
  && D_l^{(s)} = \frac{(-1)^l}{(2l+1)!} \sum\limits_{t=l}^{s-1}
    \frac{(-1)^t (4t+3)(2t+2l+2)!}{2^{2t+1}(2s-2t-1)(s+t+1)(t-l)!(t+l+1)!} \; ; \nonumber\\
  && \gamma=1.78107\;;\;\; Ei(x) \;\mbox{-- the exponential integral function [1];}\;\;
     \nonumber \\
  && \xi = \frac{\alpha+\beta}{|\vec\alpha-\vec\beta |} \;,\;\; \eta =
    \frac{\alpha-\beta}{|\vec\alpha-\vec\beta |} \;. \nonumber
\end{eqnarray}

\medskip

The matrix element of interaction of  electron with the potential center with
functions of different centers is:
\begin{eqnarray}
  && u \!\left( \vec\alpha,\vec\beta \right) =
  \int \frac{d\vec{r}}{r} \,
    \psi_k \!\left( \vec{r}-\vec\alpha \right) \,
    \psi_k \!\left( \vec{r}-\vec\beta \right)\\
  && = Z \frac{\rho^{2k+2}}{(2k+2)!} \sum\limits_{s=0}^{k+1}
    (4s+1) P_{2s}(\eta)
    \Biggl\{
    Q_{2s}(\eta) \sum\limits_{n=0}^{2k+2}
    \left[e^{-\rho} \rho^n - e^{-\rho\xi}(\rho\xi)^n \right]
    F_{s,n}^{(k+1)}(\rho)  \nonumber\\
  && \quad
    -P_{2s}(\xi) \frac{e^{-\rho\xi}}{\rho^2}
    \tilde{\tilde\Sigma}_{s,0}^{(k+1)}(x,\xi) +
    \frac{P_{2s}(\xi)}{2} \sum\limits_{i=0}^{k+1-s} a_i^{(k+1,s)}
    \Biggl[
    \ln\frac{\xi+1}{\xi-1} \cdot \frac{e^{-\rho\xi}}{\rho}
    \sum\limits_{l=0}^{2s+2i} \frac{\sigma_{2i,l}^{(s)}(\xi)}{(\rho\xi)^l}   \nonumber\\
  && \quad\quad
    +\sum\limits_{l=0}^{2s+2i} \frac{\sigma_{2i,l}^{(s)}(1)}{\rho^{l+1}} \,
    \biggl( \lefteqn{\phantom{\sum}}
    (-1)^{l+1} e^{\rho} Ei(-\rho(\xi+1)) + e^{-\rho} Ei(-\rho(\xi-1))   \nonumber\\
  && \quad\quad\quad
    -e^{-\rho\xi} \sum\limits_{h=0}^{l-1} \frac{\rho^h}{h!}
    S_{h,l} \left((-1)^{l+1}(\xi+1)^h + (\xi-1)^h \right)
    \biggr) \Biggr] \Biggr\}  \;. \nonumber
\end{eqnarray}

\newpage

\begin{center}
{\Large Tables}
\end{center}

Table 1. The results of numerical minimization of the full energy functional
for the $1s^2$-configuration ($k=0$) at $Z=Z_0$.

\begin{tabular}{|c|c|c|c|c|}
\hline $\quad Z \quad$ & $\quad \alpha \quad$ & $\quad \beta \quad$ & $\quad
\varphi \quad$ & $\quad E(\vec{\alpha},\vec{\beta}) \quad$ \\ \hline $1.5$ &
$0.078$ & $0.078$ & $3.1415$ & $-1.3140$ \\ $1.6$ & $0.068$ & $0.068$ &
$3.1415$ & $-1.5614$ \\ $1.7$ & $0.059$ & $0.059$ & $3.1415$ & $-1.8287$ \\
$1.8$ & $0.051$ & $0.051$ & $3.1415$ & $-2.116$ \\ $1.9$ & $0.046$ & $0.046$ &
$3.1415$ & $-2.4236$ \\ $2.0$ & $0.043$ & $0.043$ & $3.1415$ & $-2.7510$ \\
$2.1$ & $0.041$ & $0.041$ & $3.1415$ & $-3.0984$ \\ $2.2$ & $0.038$ & $0.038$ &
$3.1415$ & $-3.4658$ \\ $2.3$ & $0.036$ & $0.036$ & $3.1415$ & $-3.8533$ \\
$2.4$ & $0.034$ & $0.034$ & $3.1415$ & $-4.2608$ \\ $2.5$ & $0.033$ & $0.033$ &
$3.1415$ & $-4.6882$ \\ \hline
\end{tabular}

Table 2. The results of numerical minimization of the full energy functional
for the $1s^2$-configuration ($k=0$) at $Z_0=2.0$.

\begin{tabular}{|c|c|c|c|c|}
\hline $\quad Z \quad$ & $\quad \alpha \quad$ & $\quad \beta \quad$ & $\quad
\varphi \quad$ & $\quad E(\vec{\alpha},\vec{\beta}) \quad$ \\ \hline $1.5$ &
$0.0$ & $0.0$ & $-$ & $-2.8125$ \\ $1.6$ & $0.0$ & $0.0$ & $-$ & $-2.8400$ \\
$1.7$ & $0.0$ & $0.0$ & $-$ & $-2.8475$ \\ $1.8$ & $0.0$ & $0.0$ & $-$ &
$-2.8350$ \\ $1.9$ & $0.011$ & $0.011$ & $3.1415$ & $-2.8026$ \\ $2.0$ &
$0.043$ & $0.043$ & $3.1415$ & $-2.7510$ \\ $2.1$ & $0.065$ & $0.065$ &
$3.1415$ & $-2.6809$ \\ $2.2$ & $0.084$ & $0.084$ & $3.1415$ & $-2.5931$ \\
$2.3$ & $0.102$ & $0.102$ & $3.1415$ & $-2.4877$ \\ $2.4$ & $0.118$ & $0.118$ &
$3.1415$ & $-2.3650$ \\ $2.5$ & $0.132$ & $0.132$ & $3.1415$ & $-2.2253$ \\
\hline
\end{tabular}

\newpage

Table 3.

$q$-expansion of the matric elements up to $q^2$.

\begin{tabular}{ll}
\hline General expression & $q$-expansion up to $q^2$
\\
\hline
  $ S \left(\vec{q},-\vec{q}\right) \approx 1-q^2 \, 2 \int d\vec{r} \, b^2 $
& $ 1 - q^2 \frac{2 Z^2}{3(2k+1)} $
\\
\hline
  $ t \left(\vec{q},\vec{q}\right) = -\frac12 \int d\vec{r} \, a\, \Delta \,a $
& $ \frac{Z^2}{2(2k+1)} $
\\
\hline
  $ u \left(\vec{q},\vec{q}\right) \approx \int \frac{d\vec{r}}{r} a^2 +
    q^2 \int \frac{d\vec{r}}{r} \left(b^2 +a \, c\right) $
& $ Z - q^2 \frac{2Z^3}{3} \, , \quad k=0 $
\\
& $ \frac{Z}{k+1} \, , \quad k\neq 0 $
\\
\hline
  $ t \left(\vec{q},-\vec{q}\right) \approx -\frac12
    \left[ \int d\vec{r} a \, \Delta \, a + q^2 \int d\vec{r} c \, \Delta \, a \right] $
& $ \frac{Z^{2}}{2} - q^2 \frac{5Z^4}{3} \, , \quad k=0 $
\\
& $ \frac{Z^2}{2(2k+1)} - q^2 \frac{Z^4}{4k^2-1} \, , \quad k \neq 0 $
\\
\hline
  $ u \left(\vec{q},-\vec{q}\right) \approx \int \frac{d\vec{r}}{r} a^{2} +
    q^2 \int \frac{d\vec{r}}{r} \left( -b^2 + a \, c \right) $
& $ Z - q^2 \frac{4Z^3}{3} \, , \quad k=0 $
\\
& $ \frac{Z}{k+1} - q^2 \frac{2Z^3}{3(2k+1)(k+1)} \, ,\quad k \neq 0 $
\\
\hline
  $ c \left(\vec{q},-\vec{q}\right) \approx
    \int \frac{d\vec{r}_1 d\vec{r}_2}{r_{12}} a_1^2 a_2^2 + $
& $ Z \left( \frac{1}{k+1} - \frac{(4k+3)!}{2^{4k+2}
    \left[(2k+2)!\right]^2} \right) - $
\\
  $ \quad
    + q^2 \, 2 \int \frac{d\vec{r}_1 d\vec{r}_2}{r_{12}}
    \left( -2 a_1 b_1 a_2 b_2 + a_1^2 b_2^2 + a_1^2 a_2 c_2 \right)$
& $ \quad
    - q^2 Z^3 \frac{(4k+2)!}{3 \cdot 2^{4k-1} \left[(2k+2)!\right]^2} $
\\
\hline
  $ a \left(\vec{q},-\vec{q}\right) \approx
    \int \frac{d\vec{r}_1 d\vec{r}_2}{r_{12}} a_1^2 a_2^2 + $
& $ Z \left( \frac{1}{k+1} - \frac{(4k+3)!}{2^{4k+2}
    \left[(2k+2)!\right]^2} \right) + $
\\
  $ \quad
    + q^2 \, 2 \int \frac{d\vec{r}_1 d\vec{r}_2}{r_{12}}
    \left( -a_1^2 b_2^2 + a_1^2 a_2 c_2 \right) $
& $ \quad
    + q^2 Z^3 \left( -\frac{4}{3(2k+1)(k+1)}
    + \frac{(4k+2)!}{3 \cdot 2^{4k-1} \left[(2k+2)!\right]^2} \right) $
\\
\hline
  $ E^{(0)} = -\int d\vec{r} a \Delta a -
    2 Z_0 \int \frac{d\vec{r}}{r} a^2  + $
& $ Z^2 - 2Z_0 Z + \frac58 Z \, , \quad k=0 $
\\
  $ \quad
    + \int \frac{d\vec{r}_1 d\vec{r}_2}{r_{12}} a_1^2 a_2^2 $
& $ \frac{Z^2}{2k+1} - \frac{2ZZ_0}{k+1} + $
\\
& $ \quad
    + Z \left( \frac{1}{k+1} -
    \frac{(4k+3)!}{2^{4k+2} \left[(2k+2)!\right]^2} \right) \, , \quad k\neq 0 $
\\
\hline
  $ E^{(2)} = -\int d\vec{r} c \, \Delta \, a -
    \int d\vec{r} a \, \Delta \, a \int d\vec{r} b^2 - $
& $ -\frac{4Z^3}{3} \left( Z - Z_0 + \frac{3}{16} \right) \, , \quad k=0 $
\\
  $ \quad
    - 2 Z_0 \left( \int \frac{d\vec{r}}{r} a \, c +
    \int \frac{d\vec{r}}{r} a^2 \int d\vec{r} b^2  \right) + $
& $ -Z^4 \frac{4(k+1)}{3\left(4k^2-1\right)} - $
\\
  $ \quad
    + 2 \int \frac{d\vec{r}_1 d\vec{r}_2}{r_{12}}
    \left[ a_1^2 a_2 c_2 - a_1 b_1 a_2 b_2 + a_1^2 a_2^2
    \int d\vec{r} b^2 \right] $
& $ \quad
    - Z^3 \frac{(4k+3)!}{3 \cdot 2^{4k+1}(2k+1)
    \left[(2k+2)!\right]^2} \, , \quad k \neq 0 $
\\
\hline
\end{tabular}

Here $r_{12} = \left| \, \vec{r}_1 - \vec{r}_2 \, \right| $, and the index $i$
of the functions $a$, $b$, $c$ indicates dependence on $\vec{r}_i$.

\newpage
Table 4.

\begin{tabular}{|c|c|c|c|c|c|}
\hline $\, k \,$ & $E_{min}$ & $q_{min}$ & $Z_{min}$ & $E_{min}(q=0)-E_{min}$ &
$ Z_{min}(q=0)-Z_{min}$ \\ \hline
$0$ & $-2.84766$ & $0.0$ & $1.6875$ & $0$ &
$0$\\
$1$& $-2.22965$ & $0.3437$ & $2.7110$ & $0.2205$ & $-0.2559$ \\
$2$ &
$-1.79140$ & $0.5061$ & $3.2084$ & $0.3463$ & $-0.5204$ \\
$3$ & $-1.48131$ &
$0.6385$ & $3.4999$ & $0.3638$ & $-0.7030$ \\
$4$ & $-1.25036$ & $0.7644$ &
$3.6778$ & $0.3424$ & $-0.8192$ \\ \hline
\end{tabular}

\newpage

\begin{center}
{\Large Figures}
\end{center}

\bigskip

\bigskip

\begin{figure}[th]
\begin{center}
\includegraphics[width=0.7\linewidth]{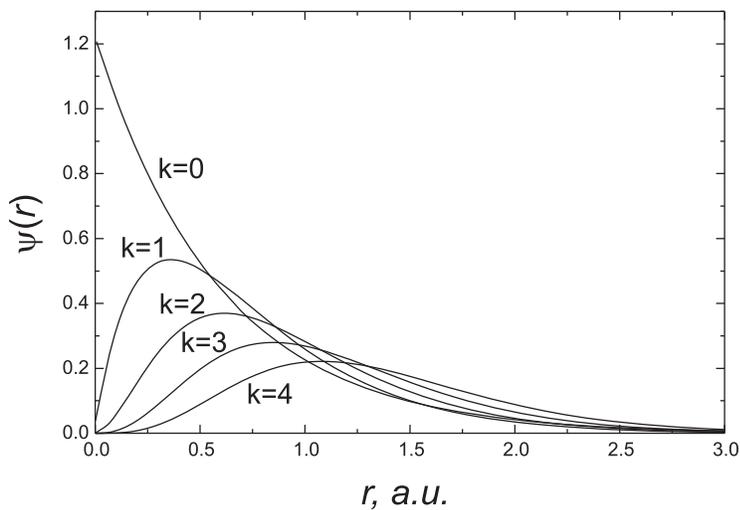}
\caption{The Slater orbitals $\psi(r)$ (\ref{WFSlat}) at $Z=Z^{(k)}_{min}$.}
\label{orbitals}
\end{center}
\end{figure}

\bigskip

\begin{figure}[th]
\begin{center}
\includegraphics[width=0.7\linewidth]{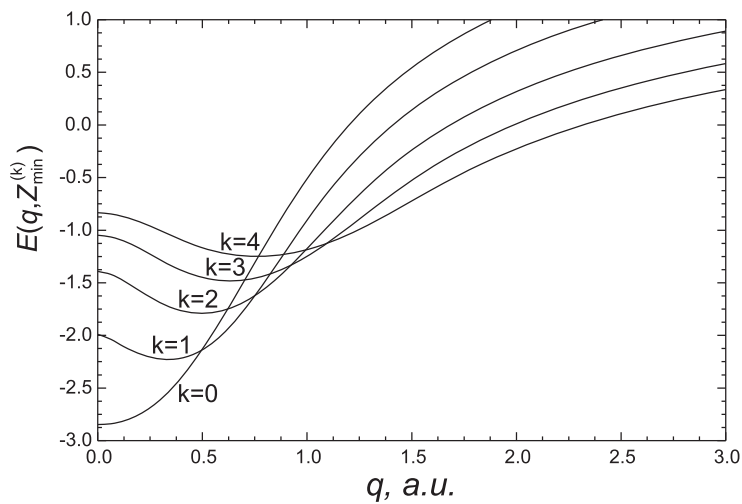}
\caption{The full energy functional $E(q,Z)$ at $Z=Z^{(k)}_{min}$.}
\label{energyfunc}
\end{center}
\end{figure}

\bigskip

\begin{figure}[thb]
\begin{center}
\leavevmode
\includegraphics[width=0.7\linewidth]{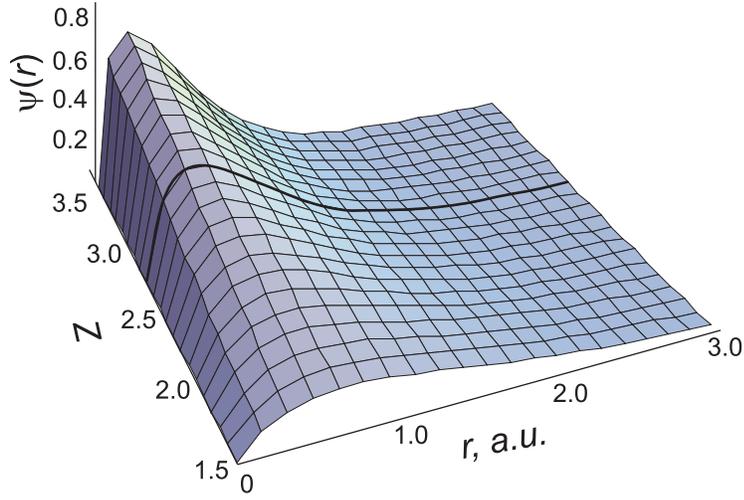}
\caption{The variation of the form of the one-particle function $\psi(r)$ at
$k=1$ with variation in $Z$. The bold line corresponds to $Z=Z^{(1)}_{min}$.}
\label{wavefunc}
\end{center}
\end{figure}

\bigskip

\begin{figure}[thb]
\begin{center}
\leavevmode
\includegraphics[width=0.7\linewidth]{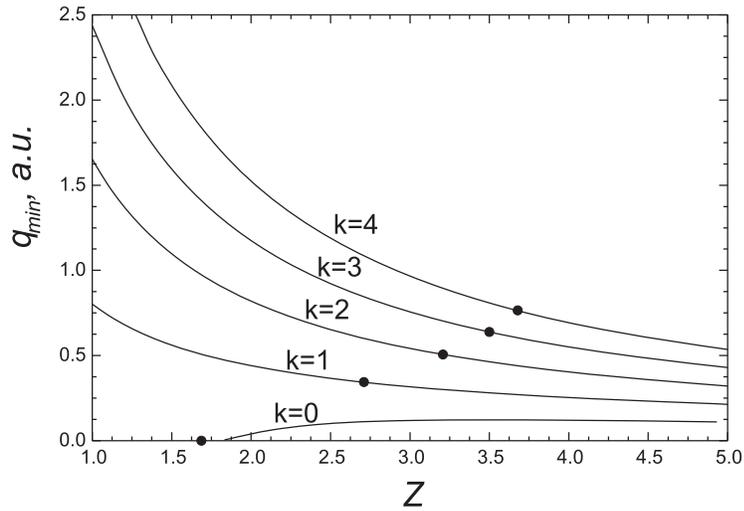}
\caption{The value of the electronic shell displacement $q$ minimizing the full
energy as function of the given parameter $Z$. The points correspond to minimal
values of the full energy at the given $k$.}
\label{shelldisp}
\end{center}
\end{figure}

\bigskip

\begin{figure}[thb]
\begin{center}
\leavevmode
\includegraphics[width=0.7\linewidth]{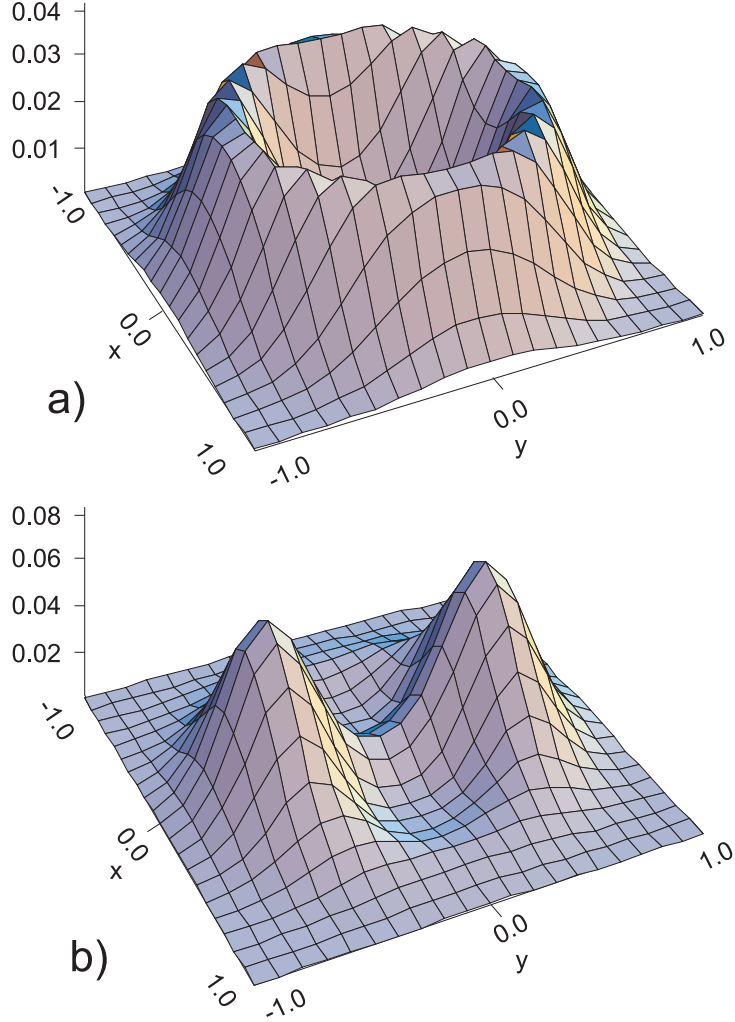}
\caption{The two-electron density distribution in plane $z=0$ for the state
$\Psi \left( \vec{r}_1,\vec{r}_2 ; \vec{q} ,-\vec{q} \right)$,  $k=1$, $Z_0=2$
a) without shifts of the electronic shells ($Z=Z^{(1)}_{min}$ at $\vec{q}=0$ );
b) with shifts $Z=Z^{(1)}_{min}$, $\vec{q}=(q^{(1)}_{min},0,0)$.}
\label{densdistr}
\end{center}
\end{figure}

\end{document}